\documentclass[fleqn,usenatbib,twocolumn]{mnras}

\usepackage{newtxtext,newtxmath}
\usepackage{xspace}
\usepackage[T1]{fontenc}

\DeclareRobustCommand{\VAN}[3]{#2}
\let\VANthebibliography\thebibliography
\def\thebibliography{\DeclareRobustCommand{\VAN}[3]{##3}\VANthebibliography}

\usepackage{graphicx}	
\usepackage{amsmath}	

\usepackage[version=3]{mhchem}
\usepackage{booktabs}
\usepackage{multirow}
\usepackage{comment}
\usepackage{physics} 
\usepackage{xcolor}

\newcommand{\mum}{$\rm \mu \mathrm{m}~$}

\newcommand{\OIII}{[\textsc{Oiii}]\,}

\newcommand{\HII}{H{\sc ii}~}

\usepackage{color}

\newcommand{\notice}[1]{\textcolor{orange}{\bf #1}}
\def\gsim{\lower.5ex\hbox{\gtsima}} 
\def\lsim{\lower.5ex\hbox{\ltsima}} 
\def\gtsima{$\; \buildrel > \over \sim \;$} 
\def\ltsima{$\; \buildrel < \over \sim \;$} \def\gsim{\lower.5ex\hbox{\gtsima}} 
\def\lsim{\lower.5ex\hbox{\ltsima}} 
\def\simgt{\lower.5ex\hbox{\gtsima}} 
\def\simlt{\lower.5ex\hbox{\ltsima}}

\newcommand{\figdir}{.}

\title[Dusty outflows from early galaxies]{Radiation-driven dusty outflows from early galaxies}

 \author[Y.Nakazato and A.Ferrara]{
 Yurina Nakazato$^{1, 2}$\thanks{Contact e-mail: \href{mailto:yurina.nakazato@phys.s.u-tokyo.ac.jp}{yurina.nakazato@phys.s.u-tokyo.ac.jp}},
 Andrea Ferrara$^{3}$,
 \\
$^{1}$ Department of Physics, School of Science, The University of Tokyo, Bunkyo, Tokyo 113-0033, Japan \\
$^{2}$ Center for Computational Astrophysics, Flatiron Institute, 162 5th Avenue, New York, NY 10010\\
$^{3}$ Scuola Normale Superiore, Piazza dei Cavalieri 7, 56126 Pisa, Italy
}

\begin{document}
\label{firstpage}
\pagerange{\pageref{firstpage}--\pageref{lastpage}}
\maketitle

 \begin{abstract}
The \textit{James Webb Space Telescope} (JWST) has discovered an overabundance of UV-bright ($M_{\rm UV} \lesssim -20$), massive galaxies at $z \gtrsim 10$ in comparison to pre-JWST theoretical predictions. Among the proposed interpretations, such excess has been explained by negligible dust attenuation conditions following radiation-driven outflows launched by young stars when a galaxy goes through a super-Eddington phase.  Dust opacity decreases the classical Eddington luminosity by a (boost) factor $A$, thus favoring the driving of outflows by stellar radiation in compact, initially dusty galaxies. Here, we compute $A$ as a function of the galaxy stellar mass, gas fraction, galaxy size, and metallicity (a total of 8 parameters). We find that the main dependence is on metallicity and, for the fiducial model, $A \sim 1800(Z/Z_\odot)/(1+N_{\rm H}/10^{23.5}\, {\rm cm^2})$. We apply such results to 20 spectroscopically confirmed galaxies at $z \gtrsim 10$ and evaluate their modified Eddington ratio. We predict that three galaxies are in the outflow phase. Their outflows have relatively low velocities ($60 -100 \,{\rm km\ s^{-1}}$), implying that they are unlikely to escape from the system. For the remaining 17 galaxies that are not currently in the outflow phase, we calculate the past evolution of the modified Eddington ratio from their star formation history. We find that 15 of them experienced an outflow phase prior to observation during which they effectively displaced their dust to larger radii. Thus, dusty outflows driven by stellar radiation appear to contribute to the observed bright UV galaxies at $z > 10$.
\end{abstract}




\section{Introduction} \label{sec:intro}
The James Webb Space Telescope (JWST) has revolutionized our understanding of the early universe, particularly by probing galaxies at $z > 10$ \citep[e.g.][]{Carniani:2024, Hainline:2024, Wang:2023, Witstok:2024, D'Eugenio:2023, Castellano:2024, Tacchella:2023, Robertson:2023, Atek:2023, Hsiao:2023, Goulding:2023, Arraba_Haro:2023_ApJL, Arrabal_Haro:2023}. One of the key JWST discoveries is the unexpected abundance of UV-bright, massive galaxies. In fact, multiple observational measurements of the UV luminosity function (LF) at $z \gtrsim 10$ find a significantly larger number of bright galaxies than predicted by pre-JWST theoretical models \citep{Harikane:2024, Harikane:2023, Finkelstein:2023, Donnan:2024, Perez-Gonzalez:2023, McLeod:2024, Yan:2023}. 

Several solutions have been proposed to explain this tension. One possibility is a top-heavy IMF which increases the luminosity-to-stellar mass ratio compared to a bottom-heavy IMF \citep{Inayoshi:2022, Chon:2022, Schaerer:2024, Yung:2024}. Another suggestion is that galaxies could form the majority of their stars within a short timescale before the feedback becomes effective, resulting in high star formation efficiency that can account for the observed UVLF \citep{Dekel:2023, Li:2023_FFB, Ceverino:2024, Shuntov:2024}. Stochastic bursty star formation histories (SFHs) can also flatten the bright-end of the UV LF \citep{Mason:2023, Mirocha:2023, Pallottini:2023, Shen:2023, Sun:2023, Gelli:2024, Jeong:2024}.

Furthermore, modified $\Lambda$CDM models may alleviate the tension. Tilting the primordial power spectrum more ‘blue’ at scales $k \gtrsim 1\, h\,{\rm cMpc^{-1}}$, which are not constrained by the cosmic microwave background (CMB), can increase the abundance of massive halos at $z>9$ \citep[e.g.][]{Hirano:2024, Parashari:2023, Ralegankar:2024}. Some studies also propose that the presence of primordial black holes (PBHs) could explain the excess of massive halos \citep{Liu:2023_PBH, Colazo:2024, Huang:2024, Zhang:2024}.

Another potential solution involves dusty outflows that can displace dust on larger radii, thus decreasing the dust optical depth at the same time, making galaxies brighter and bluer  \citep{Ferrara:2023, Ferrara:2024_z14,Tsuna:2023, Fiore:2023, Ziparo:2023}.  These outflows are likely driven by the powerful radiation field produced by the young stellar population of these compact objects, exerting a strong radiation pressure on the coupled dust-gas interstellar medium. This process resembles the analog one at work in quasar outflows, which results from the super-Eddington luminosity of the central source such as an accreting black hole for quasars and local star-forming regions \citep[e.g.][]{Murray:2005, Blackstone:2023}.

While the Eddington ratio has traditionally been used as a criterion for radiation-driven outflows, the classical Eddington luminosity assumes pure hydrogen gas, an optically thin medium to Thomson scattering, and considers only gravity from the central source. However, the vigorous star formation in compact galaxies at $z > 10$  produces considerable amounts of dust, whose absorption and scattering of UV photons must be taken into account. Also, recent JWST observations report high gas column densities of high-redshift galaxies \citep[e.g.][]{Heintz:2024, Hainline:2024, D'Eugenio:2023}, implying that the gravity of the surrounding gas cannot be ignored. 

To account for these factors, the concept of a modified Eddington luminosity has been introduced, which incorporates dust opacity effects. \citet{Fabian:2006} showed that the radiation pressure on a dusty gas can decrease the Eddington luminosity, thereby boosting the Eddington ratio. This boost factor has primarily been calculated in the context of black hole feedback from AGN \citep{Fabian:2006, Fabian:2008, Arakawa:2022, Ishibashi:2015}. 

In this study, we extend this approach by calculating the modified Eddington luminosity under conditions that cover the physical properties of compact galaxies at $z >10$. We derive the corresponding boost factor $A$ to quantify the increase in the Eddington ratio. Finally, we apply this boost factor to observed galaxies at $z > 10$.

\begin{figure}
    \centering
    \includegraphics[width = \linewidth, clip]{\figdir/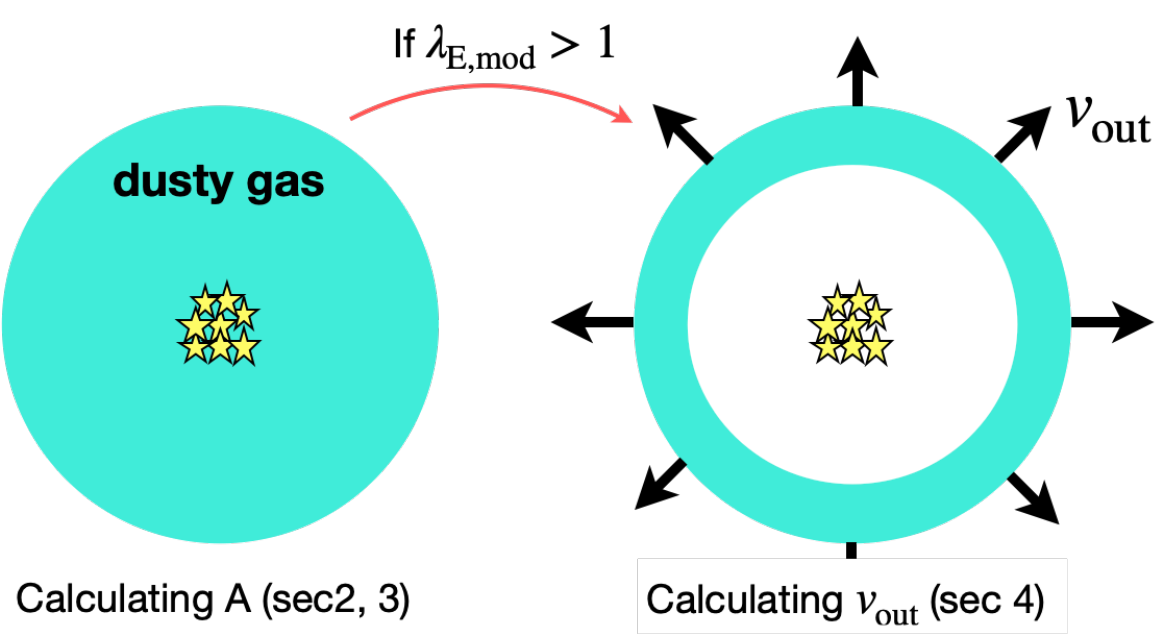}
    \vspace{-0.6cm}
    \caption{
    Schematic illustration of our modeling approach. We assume a spherically distributed gas configuration prior to the outflow (left panel), and calculate the boost factor $A$ to evaluate whether radiation pressure can overcome gravity. For systems with a modified Eddington ratio $\lambda_{\rm E,mod} > 1$ (or above the threshold value $\lambda_{\rm thres}$), we assume they enter the outflow phase and compute the corresponding outflow velocity (right panel).
}
    \label{fig:schematic_picture}
\end{figure}

The aim of this paper is to investigate how the boost factor $A$ depends on key physical parameters, and to test whether dusty outflows can be driven in observed high-redshift galaxies. The outline of the paper is as follows. Section \ref{sec:method} introduces the analytical framework of the modified Eddington luminosity and defines the key parameters used in the model. In Section \ref{sec:result}, we compute the boost factor $A$ for a uniform gas sphere, exploring its dependence on various physical properties such as stellar and gas masses, galaxy size, dust composition, stellar age, and metallicity. We identify the most influential parameters and provide an analytical formula for $A$ as a function of these quantities. In Section \ref{sec:application_obsz10}, we apply this analysis to observed galaxies at $z > 10$ to examine the onset of a dusty outflow. For galaxies where an ongoing outflow is expected, we estimate the outflow velocity using a shell-like geometry. Figure \ref{fig:schematic_picture} illustrates our modeling framework in this study. For galaxies not currently showing outflows, we further investigate whether a past outflow phase is plausible by assuming delayed-tau star formation histories. Finally, Section \ref{sec:conclusion} summarizes our main results.

\section{Method} \label{sec:method}
\subsection{Basic physical arguments} \label{subsec:analytical_configration}
\subsubsection{Classical Eddington Luminosity}
The classical Eddington luminosity, $L_{\rm E}$, is derived assuming spherical geometry, a pure ionized hydrogen gas that is optically thin to Thomson scattering; gravity is from the central source only, i.e., stars in our case. Under these assumptions, we get: 
\begin{equation}
    L_{\rm E} = \frac{4\pi GM_*m_{\rm p}c}{\sigma_{\rm T}}, \label{eq:Eddington}
\end{equation}
where $m_{\rm p}$ is the proton mass, and $\sigma_{\rm T} = 6.65 \times 10^{-25}\, {\rm cm^{-2}}$  is the Thomson scattering cross section. Other symbols have the usual meaning. The Eddington ratio is defined as $\lambda_{\rm E} = L / L_{\rm E}$, where $L$  is the bolometric luminosity of the galaxy. 

\subsubsection{Modified Eddington Luminosity}\label{subsubsec:modified_eddington_luminosity}
Here we assume spherical geometry, ionized hydrogen gas, and dust, and consider the gravity of the central luminous source and the surrounding gas as an initial condition. Dust and gas are tightly coupled by viscous and Coulomb drag forces. Dust grains in the gas interact with photons of wavelength $\lambda$ with the radiation pressure cross section per H nucleon $\sigma_{{\rm pr,d}}(\lambda)\, [{\rm cm^2/H}]$. The radiative force per proton at radius $r$ can be expressed as
\begin{equation}
    F_{{\rm R}, \lambda} = \frac{L_{\lambda}(1-e^{-\tau_{{\rm ext}}(\lambda)})\sigma_{{\rm pr}}(\lambda)}{4\pi r^2 c}, \label{eq:radiative_force_per_wavelength}
\end{equation}
where $\sigma_{\rm pr}$ is the wavelength-dependent radiation pressure cross section including both dust radiation pressure and Thomson scattering, $\sigma_{\rm pr}(\lambda) = \sigma_{\rm pr,d}(\lambda) + \sigma_{\rm T}$.  In turn, $\sigma_{\rm pr}$ depends on the absorption and scattering cross sections per H nucleus and on the asymmetry parameter $\langle \cos \theta \rangle$  \citep{Weingartner:2001}: 
\begin{equation}
    \sigma_{\rm pr, d}(\lambda) = \sigma_{\rm abs, d}(\lambda) + (1- \langle \cos \theta \rangle_\lambda)\sigma_{\rm sca, d}(\lambda).
\end{equation} 
The dust absorption and scattering cross sections are linearly scaled with the dust-to-gas ratio normalized to the adopted extinction curve, i.e., Milky-Way or SMC, $\sigma_{i,{\rm d}} \propto \mathcal{D}/\mathcal{D_{\rm{MW, SMC}}}$, where $i= (\rm abs, sca)$, and  $\mathcal{D}_{\rm MW} = 1/165$,  $\mathcal{D}_{\rm SMC}= 1/667$.  Observations also show \citep[e.g.][]{Remy-Ruyer:2014,Birkin:2014} that the dust-to-gas ratio scales with metallicity as $\mathcal{D} \propto(Z/Z_\odot)$. We assume that gas metallicity is the same as the stellar one. The exact nature of dust grains in high-redshift galaxies is a matter of ongoing debate. Recent JWST observations of galaxies at $z > 4$ \citep{Fisher:2025, Ormerod:2025, Witstok:2023, Markov:2023, Markov:2025} show that 20–25\% of the REBELS and JADES samples exhibit evidence for a 2175 \AA\, bump, likely originating from PAHs \citep{Li:2024, Lin:2025, Nanni:2025}, as seen in the MW extinction curve \citep{Cardelli:1989} but not in the SMC curve \citep{Gordon:2003}. We therefore adopt these two models for simplicity, with the MW dust type taken as our fiducial case.

The cumulative extinction optical depth is defined as $\tau_{\rm ext} (\lambda)\equiv [\sigma_{\rm ext,d}(\lambda) + \sigma_{\rm T}] N_{\rm H}$, where $\sigma_{\rm ext, d}$ and $N_{\rm H}$ are dust extinction cross section, and the gas column density, respectively. 

By integrating eq. (\ref{eq:radiative_force_per_wavelength}) over the wavelength, we obtain 
\begin{equation}
    F_{\rm R} = \int F_{{\rm R}}(\lambda) {\rm d}\lambda \simeq \frac{L_{\rm bol}(1-e^{-\tau_{\rm ext}})\sigma_{\rm pr}}{4\pi r^2 c} \label{eq:F_R}.
\end{equation}
Since the dust cross section depends on wavelength, we have defined the spectrum-averaged cross section $\sigma_{\rm pr}, \, \sigma_{\rm ext}$ by weighting it with the SED shape:
\begin{align}
    \sigma_{\rm pr} &= \frac{\int \sigma_{\rm pr, d}(\lambda) L_{\lambda}d\lambda}{\int L_\lambda d\lambda}  + \sigma_{\rm T}, \quad \sigma_{\rm ext} = \frac{\int \sigma_{\rm ext, d}(\lambda) L_{\lambda}d\lambda}{\int L_\lambda d\lambda} + \sigma_{\rm T}, \label{eq:sigma_ave}
\end{align}

The gravitational force on a proton originates from the central stars and the surrounding gas;
\begin{equation}
    F_{\rm G} = F_{\rm G, star} + F_{\rm G,  gas}
              = \frac{GM_* m_{\rm p}}{r^2} + F_{\rm G,  gas}. \label{eq:F_G}
\end{equation}
The gas gravity varies depending on the gas distribution. Here we consider two density distributions: (i) a uniform sphere, and (ii) a shell. Note that in our setup we consider a uniform gas distribution only within the central $\sim$100 pc region, motivated by the observed compactness of $z > 10$ galaxies. Such compact systems resemble dusty outflows around \HII regions \citep[10-40 pc scale;][]{Blackstone:2023, Pathak:2025} rather than halo-scale AGN-driven winds \citep[e.g.,][]{Costa:2018_cosmological, Thompson:2015}. This choice allows for a simple analytic treatment while still capturing the key physics relevant for the onset of outflows.

Then we derive the modified Eddington luminosity at $F_{\rm R} = F_{\rm G}$ using eq. (\ref{eq:F_R}) and (\ref{eq:F_G});
\begin{align}
    L_{\rm E, mod} &= \frac{4\pi r^2 c}{(1-e^{-\tau_{\rm ext}})\sigma_{\rm pr}} \label{eq:modified_Eddington}\, F_{\rm G}\notag \\
    &= \left( \frac{GM_* m_{\rm p}}{r^2} + F_{\rm G,  gas}\right)\, \frac{4\pi r^2 c}{(1-e^{-\tau_{\rm ext}})\sigma_{\rm pr}}, 
\end{align}
where
\begin{align}
    F_{\rm G, gas} &= \frac{4\pi Gm_{\rm p}^2 rn_{\rm H}}{3} =\frac{4\pi G m_{\rm p}^2 N_{\rm H}}{3}\quad (\text{uniform})\\
    F_{\rm G, gas} &= \frac{4\pi}{3} \left(r^3 - r_{\rm in}^3\right)n_{\rm H} m_{\rm p}\, \frac{Gm_{\rm p}}{r^2} \quad (\text{shell})
\end{align}
The boost factor is defined as $A\equiv {L_{\rm E}}/L_{\rm E, mod}$. From this definition, and using eq. (\ref{eq:Eddington}) and eq. (\ref{eq:modified_Eddington}), it follows that 
\begin{equation}
	A\equiv \frac{L_{\rm E}}{L_{\rm E, mod}} =\frac{F_{\rm G,star}}{F_{\rm G}} \,\frac{(1-e^{-\tau_{\rm ext}})\sigma_{\rm pr}}{\sigma_{\rm T}}. \label{eq:boost_factor}
\end{equation}

Since our primary goal is to investigate the onset of outflows, we include the contribution of gas gravity. The boost factor $A$ calculated here represents the expectation prior to an outflow being generated. After the onset, we adopt the shell model to calculate the outflow velocity (Section \ref{sec:application_obsz10}). We neglect the contributions of the dark matter halo and central black holes because the compact central regions we consider are dominated by baryonic gravitational potentials. Moreover, for galaxies with stellar masses of about $10^8\, M_\odot$, the corresponding black hole mass is expected to be only $\sim 10^6\, M_\odot$ assuming $M_{\rm BH}/M_* \sim 0.01$, as recently derived from high-redshift ($z\sim3-7$) AGN observations with JWST \citep{Pacucci:2023, Jones:2025}. The estimated BH mass is too small to contribute significantly as a gravitational source or additional photon source.

\subsection{Parameter setup}
In order to calculate $A$, we set several parameters for the following three components: stellar, gas, and dust. We assume that stars are located in the central region and have total mass $M_*$. The gas mass is obtained by assigning the gas fraction $f_{\rm gas}\equiv M_{\rm gas}/(M_{\rm gas} + M_*)$; such mass is distributed within the effective radius ($r_{\rm e}$) according to one of the two prescribed spatial distributions (\texttt{distribution}), from which we obtain the gas number density $n_{\rm H}$. For the shell geometry, we assume an inner radius of $r_{\rm in} = 0.7 (0.1) \, {\rm pc}$. The inner radius is chosen because the outflow geometry is assumed to be shell-like. More details are provided in Section \ref{subsec:transition_regime}.

For the dust component, we assume Milky Way (MW) or Small Magellanic Cloud (SMC)-like dust (\texttt{dust type}). Regarding stellar SEDs, we employ BPASS SED v2.3 with single stellar evolution\footnote{According to \citet{Eldridge:2017}, BPASS single SEDs are similar to other ones such as Starburst 99 \citep{Leitherer:1999}, GALAXEV \citep[BC03;][]{Bruzual_Charlot:2003}, and \citet{Maraston:2005} at the age of $<$ 1 Gyrs.} \citep{Eldridge:2017, Byrne:2022} and broken-power law IMF 
\begin{equation}
    N(M < M_{\rm max} )\propto \int^{M_1}_{0.1} \left(\frac{M}{M_\odot}\right)^{\alpha_1} {\rm d}M 
     + M_1^{\alpha_1}\int^{M_{\rm up}}_{M_1} \left(\frac{M}{M_\odot}\right)^{\alpha_2} {\rm d}M,
\end{equation}
where $\alpha_1 = -1.3, \alpha_2 = -2.35, M_1 = 0.5\ M_\odot, M_{\rm up} = 300\ M_\odot$. We parametrize \texttt{age} and metallicity ($Z$), assuming bursty (i.e., instantaneous) star formation.

Table \ref{table:parameter} shows the parameters we set to obtain the boost factor $A$. The third column represents the fiducial values. The values of stellar mass, effective radius, age, and metallicity are typical ones of observed $z >10$ galaxies \citep[e.g.][ see the references in Table \ref{table:z10_obs}]{Carniani:2024, Hainline:2024}. The average cross sections of the fiducial model are 
\begin{align}
    \sigma_{\rm ex} &= 1.6\times 10^{-21}\, {\rm cm^{-2}} (Z/Z_\odot) \\
    \sigma_{\rm pr} &= 1.2 \times 10^{-21}\, {\rm cm^{-2}} (Z/ Z_\odot).
\end{align}
In this study, we assume ionized gas, i.e., $N_{\rm e} \simeq N_{\rm H}$. Since the cross section of dusty gas mainly depends on the dust cross section rather than that of the Thomson cross section, the result does not change largely in the case of partially ionized gas. 
\begin{table}
\centering
\begin{tabular}{lll}
\hline
parameter         & values              & fiducial      \\\hline\hline
$M_* \,[M_\odot]$ & $10^7, 10^8,  10^9$ & $10^8$ \\
$f_{\rm gas}$ &  0.33, 0.50, 0.67, 0.75     & 0.67    \\
$r_{\rm e}$ [pc]  & 50, 100, 200, 300, 400, 500  & 100   \\
\texttt{distribution}  & uniform, shell, & uniform    \\
\texttt{dust type}& MW, SMC             & MW      \\
$\mathcal{D}$               & $\mathcal{D}_{\rm MW}(Z/Z_\odot)$  & 1/1650  \\
\texttt{age} [Myr]& 10, 30, 50, 100     & 10       \\
\multirow{2}{*}{$Z$}   & $10^{-5}, 10^{-4}, 10^{-3}, 0.002, 0.003,$ & \multirow{2}{*}{0.002} \\
                 & $0.004, 0.006, 0.008, 0.01, 0.02(Z_\odot)$ & \\
\hline           
\end{tabular} 
\caption{Parameter setting. The gas fraction $f_{\rm gas}$ is defined as $f_{\rm gas} \equiv M_{\rm gas}/(M_{\rm gas}+M_*)$. We assume the gas metallicity to be equal to the stellar metallicity.}\label{table:parameter}
\end{table}

\begin{figure}
    \centering
    \includegraphics[width = \linewidth, clip]{\figdir/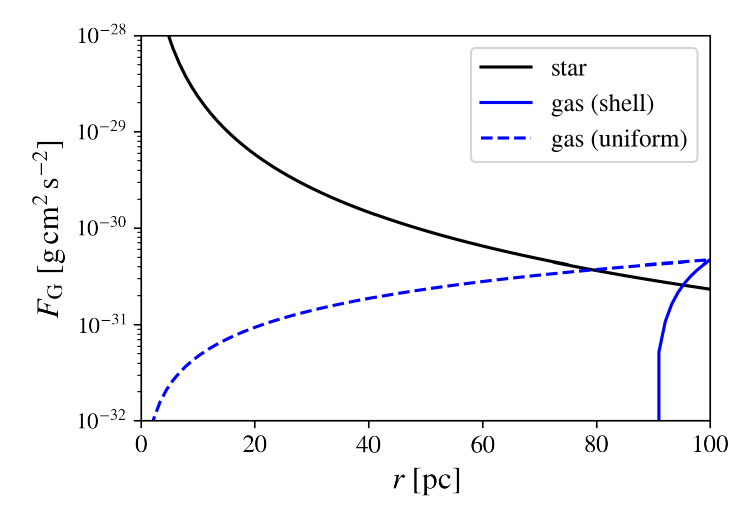}
    \vspace{-0.6cm}
    \caption{ Force acting on one proton as a function of radius. We set the fiducal values written in the Table \ref{table:parameter}; $r_{\rm e} = 100\, {\rm pc}, \, f_{\rm gas} = 2/3, M_* = 10^8 \, M_\odot$. For the shell geometry, we adopt $r_{\rm in} = 0.7 r_{\rm e}$. The black solid line shows the gravity from the central stellar components.  The blue dashed (solid) line represents the gravity from the surrounding gas with uniform (shell) geometry. 
}
    \label{fig:gas_gravity_per_proton}
\end{figure}

Figure \ref{fig:gas_gravity_per_proton} shows the gravity force acting on a proton due to the stellar and gas mass distribution as a function of radius. We adopt the fiducial value for each parameter as shown in Table \ref{table:parameter}. We see that gas gravity exceeds stellar gravity at $R \gtrsim 70 (85) \, {\rm pc}$ in the case of uniform (shell) geometry of the gas distribution. 

\section{Results} \label{sec:result}
\subsection{Radial dependence of the boost factor} \label{subsec:B_vs_radius}

\begin{figure}
    \centering
   \includegraphics[width = \linewidth, trim = 0 15 0 0, clip]{\figdir/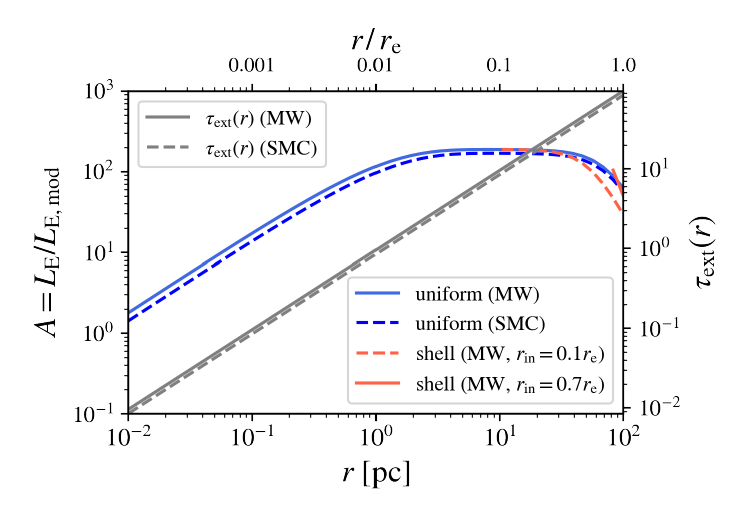}
    \vspace{-0.5cm}
    \caption{ Boost factor as a function of distance from the center. The blue solid (dashed) line represents the case of a uniform gas geometry with MW (SMC)-like dust type. The red solid and dashed lines are the cases of shell gas geometry with the same dust type (MW) but with a different inner radius of $r_{\rm in} = 0.1 r_{\rm e}$ and $0.7 r_{\rm e}$, respectively. The gray lines show the radially dependent optical depth in the uniform-sphere case (right y-axis), $\tau_{\rm ext}(r) =N_{\rm H}(r) \sigma_{\rm ext}$.
The other parameter settings ($M_*, f_{\rm gas}, r_{\rm e}, \texttt{age}, Z$) are set as the fiducial ones in Table \ref{table:parameter}. 
}
    \label{fig:A_r}
\end{figure}

Figure \ref{fig:A_r} shows the boost factor as a function of the distance from the center. We describe below the behavior in three different regimes. In this section, our aim is simply to examine the boost factor $A$ as a function of radius, using the radially dependent optical depth defined as $\tau(r) = N_{\rm H}(r)\sigma$. In the discussion of the onset of outflows, however, we adopt the optical depth at the outer radius (i.e., the effective radius $r = r_{\rm e}$), which is discussed in the next section \ref{subsec:parapeter_dependence}. 
\subsubsection{Optically thin regime}
Within the region of 1\% of the outer radius ($r_{\rm e}$), the optical depth is small and the gas mass is negligible. The gravity and radiative force are approximated as follows:
\begin{align}
    F_{\rm G} &\sim F_{\rm G, star} \label{eq:F_G_star}\\
    F_{\rm R} &\sim \frac{L\sigma_{\rm pr} \tau_{\rm ext}}{4\pi r^2 c} .
\end{align}
The modified Eddington luminosity and the boost factor become
\begin{align}
    & L_{\rm E, mod} \simeq \frac{\sigma_{\rm T}}{\sigma_{\rm pr} \tau_{\rm ext}} L_{\rm E} \\
    & A= \frac{\sigma_{\rm pr} \tau_{\rm ext}}{\sigma_{\rm T}} = \frac{\sigma_{\rm pr}\sigma_{\rm ext} N_{\rm H}}{\sigma_{\rm T}} \propto r. \label{eq:opticaly_thin}
\end{align}
\subsubsection{Transition regime} \label{subsec:transition_regime}
As the distance from the center increases to $\gtrsim 5\%$ of the outer radius ($r_{\rm e}$), the optical depth becomes $\sim$ 1, and the gas gravity is still negligible. 
The gravity is the same as eq. (\ref{eq:F_G_star}) and the radiative force is 
\begin{equation}
    F_{\rm R} \sim \frac{L\sigma_{\rm pr}}{4\pi r^2 c}. \label{eq:F_R_optically_thick}
\end{equation}
The modified Eddington luminosity and the boost factor are written as
\begin{align}
    & L_{\rm E, mod} \simeq \frac{\sigma_{\rm T}}{\sigma_{\rm pr}} L_{\rm E} \\
    & A= \frac{\sigma_{\rm pr}}{\sigma_{\rm T}} = \frac{\sigma_{\rm pr, d} + \sigma_{\rm T}}{\sigma_{\rm T}} \simeq \frac{\sigma_{\rm pr, d}}{\sigma_{\rm T}}= 1800 \left(\frac{Z}{Z_\odot}\right) .\label{eq:B_middle}
\end{align}
\subsubsection{Optically thick regime}
As the distance further increases to $\gtrsim 70\%$ of the outer radius ($r_{\rm e}$), the gas becomes optically thick ($\tau_{\rm ext} \gg 1$), and gas gravity can no longer be neglected, as shown in Figure \ref{fig:gas_gravity_per_proton}.
The radiative force is still given by eq. (\ref{eq:F_R_optically_thick}), while $F_{\rm G} \sim F_{\rm G, gas}$. The boost factor is then
\begin{align}
   A&= \frac{M_*}{M_{\rm gas} (< r)}\frac{\sigma_{\rm pr}}{\sigma_{\rm T}}  \\
   &=
   \begin{cases}
    \left[ 1 + \frac{f_{\rm gas}}{1-f_{\rm gas}} \left(\frac{r}{r_{\rm e}}\right)^3\right]^{-1}\frac{\sigma_{\rm pr}}{\sigma_{\rm T}}  \,\, (\text{uniform}) \\
     \left[ 1 + \frac{f_{\rm gas}}{1-f_{\rm gas}} \left(\frac{r^3 -r_{\rm in}^3}{r_{\rm e}^3}\right)\right]^{-1}\frac{\sigma_{\rm pr}}{\sigma_{\rm T}} \,\,(\text{shell}) .
   \end{cases}
\end{align}

Note that in this regime $A\propto r^{-3}$ for both cases. We also test a range of inner radii from $0.1r_{\rm e}$ up to $0.9r_{\rm e}$, motivated by the fact that the outflow propagates through a shell-like gas geometry. Figure \ref{fig:A_r} shows the two cases of $r_{\rm in} = 0.1 r_{\rm e}$ and $0.7 r_{\rm e}$ (fiducial). We find that the value of A in the shell case is the same as that of the uniform case at $r \lesssim 0.7r_{\rm e}$, where the gas gravity is negligible. At $r \gtrsim 0.7r_{\rm e}$, the shell geometry reduces the values of $A$ by up to a factor of three compared to the uniform case. Since the uniform geometry already encompasses the shell cases, from here onwards we concentrate on the uniform-density case.

\subsection{Parameter dependence} \label{subsec:parapeter_dependence}
\begin{figure}
    \centering
    \includegraphics[width = \linewidth, trim = 0 18 0 0, clip]{\figdir/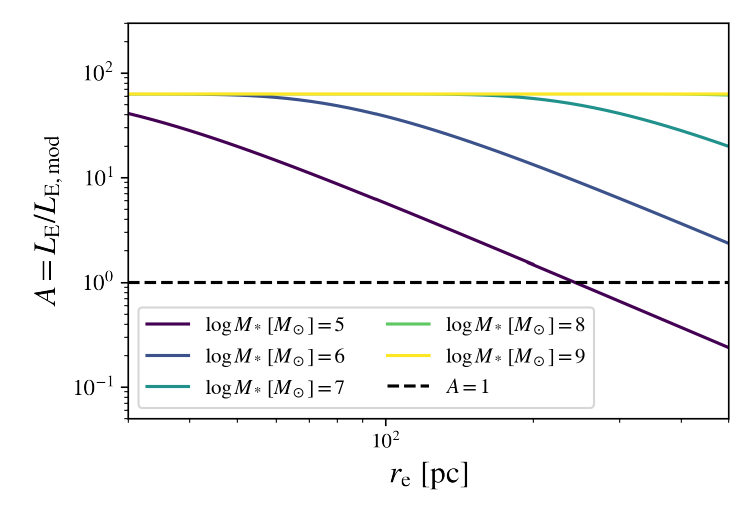}
    \vspace{-0.4cm}
    \caption{ Boost factor as a function of the galaxy effective radius. The other parameters (gas fraction, \texttt{age}, metallicity, and \texttt{dust type}) are fixed to their fiducial values shown in Table \ref{table:parameter}.
}
    \label{fig:A_re}
\end{figure}
In this Section, we investigate the dependence of the boost factor on several physical properties: the galaxy size ($r_{\rm e}$), gas fraction $f_{\rm gas}(\equiv M_{\rm gas}/(M_{\rm gas} + M_*))$, metallicity ($Z$), stellar age (\texttt{age}), and dust properties as shown in Table \ref{table:parameter}. From the following subsections, we vary one or two parameters, while keeping all others fixed at their fiducial values. From this section onwards, we evaluate the values of the boost factor $A$ at the outer radius $r_{\rm e}$, which sets the condition for gas to escape. Here we note that the boost factor $A$ is calculated directly from eq. (\ref{eq:boost_factor}), and the subsequent equations in this section are presented to provide a qualitative understanding of the trends shown in the figures.
\subsubsection{Dependence on galaxy size}
Figure \ref{fig:A_re} shows the boost factor as a function of effective radius from $r_{\rm e} = 50 \,{\rm pc}$ to 500 pc with different stellar masses ($\log_{10} M_* \,[M_\odot] = 5-9$). For the case with $M_* = 10^5\, M_\odot, \, f_{\rm gas}=0.67$, the gas mass is $M_{\rm gas} = f_{\rm gas} M_*/(1-f_{\rm gas}) = 2\times 10^5 \,M_\odot$, and the corresponding gas column density is $\sim 10^{20.5} \, {\rm cm^{-2}}(M_*/10^5 \, M_\odot)(r_{\rm e}/100 \, {\rm pc})^{-2}$. Therefore, the gas is optically thin and $A$ is calculated from eq. (\ref{eq:boost_factor}),
\begin{align}
	A &= \frac{M_*}{M_* + M_{\rm gas}} \, \frac{\tau_{\rm ext} \sigma_{\rm pr}}{\sigma_{\rm T}}  = (1-f_{\rm gas}) \, \frac{\sigma_{\rm ext}\sigma_{\rm pr} N_{\rm H}}{\sigma_{\rm T}} \notag  \\ 
 & \sim  \frac{\sigma_{\rm ext}\sigma_{\rm pr}}{\sigma_{\rm T}} \frac{f_{\rm gas} M_*}{4\pi r_{\rm e}^2} \propto r_{\rm e}^{-2}. \label{eq:B_optically_thin}
\end{align}
Here we use $N_{\rm H} \sim M_{\rm gas}/4\pi r_{\rm e}^2 = f_{\rm gas} M_*/\{4\pi r_{\rm e}^2 (1-f_{\rm gas})\}$.
As the stellar and gas masses increase, gas is optically thick and $A$ is calculated as 
\begin{equation}
A = (1-f_{\rm gas}) \, \frac{\sigma_{\rm pr}}{\sigma_{\rm T}} = {\rm const. } \label{eq:B_optically_thick}
\end{equation} 

\begin{figure}
    \centering
    \includegraphics[width = \linewidth, trim=0 12 0 0, clip]{\figdir/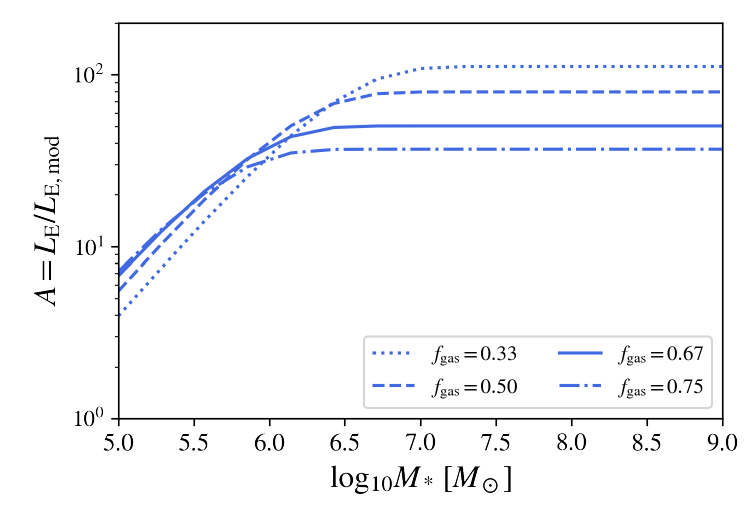}
     \vspace{-0.4cm}
    \caption{ Boost factor as a function of stellar mass. Purple, blue, and green lines represent different gas geometries. The dotted, dashed, solid, and dash-dot lines show $f_{\rm gas}=0.33, 0.50, 0.67(\text{default}), 0.75$, respectively. 
}
    \label{fig:Ms_gas_ratio}
\end{figure}

\subsubsection{Dependence on stellar mass/gas fraction}
We also check the dependence of stellar mass and the gas fraction ($f_{\rm gas}$). Figure \ref{fig:Ms_gas_ratio} shows the boost factors as a function of stellar mass with different $f_{\rm gas}$. We see that in all cases, boost factors decrease as stellar masses decrease due to a weaker radiation pressure. For  $M_* \lesssim 10^{6.0}\, M_\odot$, the boost factor $A$ is written as the same as eq. (\ref{eq:B_optically_thin}), 
\begin{equation}
   A \sim \frac{\sigma_{\rm ext}\sigma_{\rm pr}}{\sigma_{\rm T}}\frac{f_{\rm gas} M_*}{4 \pi r_{\rm e}^2} \propto f_{\rm gas} M_* \label{eq:B_optically_thin_fgas}
\end{equation}
Therefore, when the gas is optically thin, $A$ is proportional to the gas fraction and the stellar mass. When the stellar mass increases and the surrounding gas becomes optically thick, the boost factor becomes constant with respect to the stellar mass but inversely proportional to gas fraction, as shown in eq. (\ref{eq:B_optically_thick}). Figure \ref{fig:Ms_gas_ratio} shows that the dependence of $A$ on the gas fraction reverses around a stellar mass of $M_* \sim 10^{6.5}\, M_\odot$.

\subsubsection{Dependence on metallicity}
\begin{figure}
    \centering
    \includegraphics[width = \linewidth, trim = 0 18 0 0 , clip]{\figdir/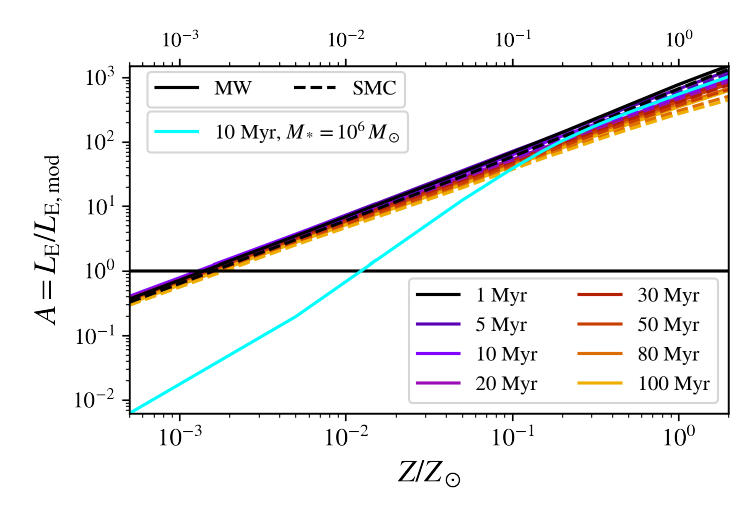}
     \vspace{-0.5cm}
    \caption{ Boost factor as a function of metallicity. Solid (dashed) lines are in the case of MW(SMC) dust. Colors represent different stellar ages. We adopt the default values of stellar mass ($M_* = 10^8 \,M_\odot$), gas fraction, radius, and stellar age in Table \ref{table:parameter}. The cyan line represents the case with a smaller stellar mass of $M_* = 10^6\, M_\odot$. 
}
    \label{fig:A_Z}
\end{figure}

We further investigate the dependence of the boost factor on the metallicity. Figure \ref{fig:A_Z} represents the boost factors as a function of metallicity with different stellar ages. 
From the above discussions, gas is optically thick for the fiducial stellar mass and gas fraction, and $A$ is written in the same way as eq. (\ref{eq:B_optically_thick}), 
\begin{equation}
   A = (1-f_{\rm gas}) \frac{\sigma_{\rm pr}}{\sigma_{\rm T}} \propto \frac{\sigma_{\rm pr, d} + \sigma_{\rm T}}{\sigma_{\rm T}} \sim \frac{\sigma_{\rm pr, d}}{\sigma_{\rm T}} \propto Z. \label{eq:A_fig5}
\end{equation}
The boost factor is proportional to metallicity, as shown in Figure \ref{fig:A_Z}. We have also plotted the case with a stellar mass of $M_* = 10^6\, M_\odot$ as a cyan solid line. In this case, gas is optically thin and $A$ is calculated as in the same as eq. (\ref{eq:B_optically_thin}) and eq. (\ref{eq:B_optically_thin_fgas}), 
\begin{equation}
   A \sim \frac{\sigma_{\rm pr}\sigma_{\rm ext}}{\sigma_{\rm T}}\frac{f_{\rm gas} M_*}{r_{\rm e}^2} \propto M_*  Z^2 .
\end{equation}

\subsubsection{Dependence on stellar age}
We investigate boost factors for different stellar ages of SEDs. As stellar age increases, the SEDs become softer \citep[e.g.][]{Xiao:2018}, leading to a smaller spectrum-averaged cross section $\sigma_{\rm pr}$ (eq. (\ref{eq:sigma_ave})). This results in weaker radiative pressure (eq. (\ref{eq:F_R})). Figure \ref{fig:A_Z} illustrates the boost factors for various stellar ages. The boost factors decrease with increasing stellar age, but the change is limited to a factor of $\lesssim 2$ for all the gas distributions considered.

\subsubsection{Dependence on dust properties}
We finally investigate the dependence on the dust properties, such as the grain size distribution and composition, resulting in different absorption and scattering efficiencies \citep[e.g.][]{Draine:2011}. We use models from \citet{Weingartner:2001} that are appropriate for MW and SMC extinction curves. As discussed in Section \ref{subsec:analytical_configration}, the physical properties of dust grains in high-redshift galaxies, such as composition and size distribution, remain uncertain \citep[e.g.][]{Markov:2023, Markov:2025}. As the dashed lines in Figure \ref{fig:A_Z} show, the boost factor $A$ for SMC-dust differs only by a factor of $\sim$1.1 with respect to MW-dust cases. This is because the spectrum-averaged cross section $\sigma_{\rm pr}$ is dominated by the value in the  UV range, i.e., $\sigma_{\rm pr} \sim \sigma_{\rm pr, UV}$, where the MW and SMC curves have a similar value, $\sigma_{\rm pr, UV (MW)}= 8.7 \times 10^{-22} \, {\rm cm^{2}} (Z/Z_\odot)$ and $\sigma_{\rm pr, UV (SMC)}= 9.2 \times 10^{-22} \, {\rm cm^{2}} (Z/Z_\odot)$, respectively.

\subsection{Boost factor as a function of $N_{\rm H}$ and $Z$}
\begin{figure}
    \centering
    \includegraphics[width = \linewidth, trim = 0 10 0 0, clip]{\figdir/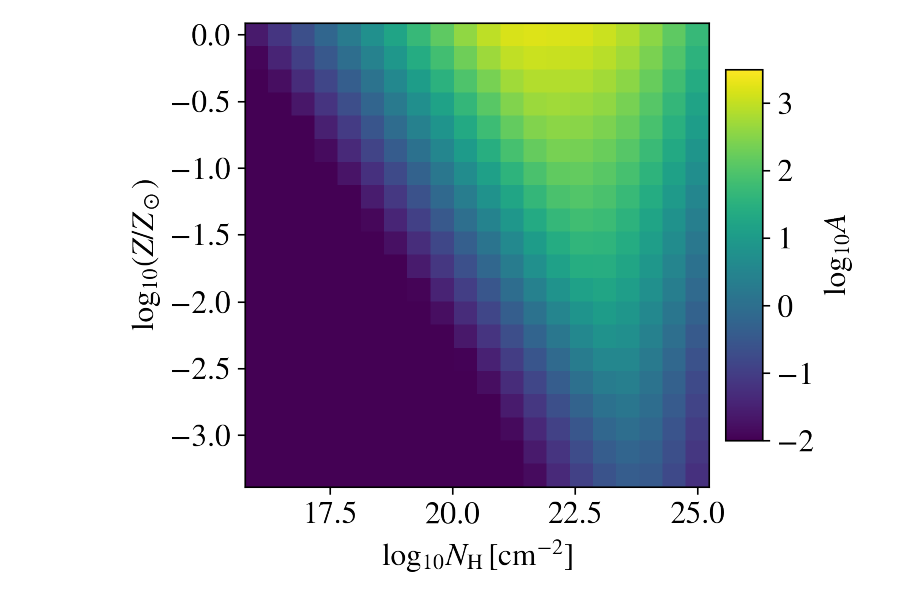}
     \vspace{-0.4cm}
    \caption{ Boost factor as a function of gas column density, $N_{\rm H}$, and metallicity, $Z$. We assume a fully ionized, pure hydrogen gas for which $N_{\rm e} = N_{\rm H}$. The other parameters are the fiducial ones in Table \ref{table:parameter}.
}
    \label{fig:A_colormap}
\end{figure}

According to section \ref{subsec:parapeter_dependence}, we find that the dependence of the boost factor on dust type and stellar age is small and can be neglected. On the other hand, the dependencies on $M_*, f_{\rm gas}$, and $r_{\rm e}$ can be consolidated into the gas column density as $N_{\rm H} \sim \int n_{\rm H}(r) {\rm d}r$ is a function of these three quantities (see eq. (\ref{eq:B_optically_thin})). Hence, we conclude that $A$ is essentially a function of $N_{\rm H}$ and $Z$ only. Such a relation is shown in  Fig. \ref{fig:A_colormap}, which can then be considered as the main results of this work. The key features of the relation can be understood as follows.

From eq. (\ref{eq:sigma_ave}) we see that the dust opacity largely dominates above the Thomson scattering one, i.e.  

\begin{align}
    \sigma_{\rm ext} &= \sigma_{\rm ext, d} + \sigma_{\rm T} \simeq \sigma_{\rm ext, UV} \notag  \\
                     &= 1.2 \times 10^{-21} \, {\rm cm^{-2}} \left(\frac{Z}{ Z_\odot}\right) .\label{eq:sigma_approx}
\end{align}
In the optically thin regime, i.e., $N_{\rm H} =10^{16-22}\, {\rm cm^{-2}}$, the gravitational contribution of gas can be neglected\footnote{We note that typical observed galaxies at $z > 10$ have stellar masses $M_* \sim 10^8\, M_\odot$ (see Table \ref{table:z10_obs}), for which the gas gravity is smaller than the stellar gravity in the optically thin regime. As shown in Figure \ref{fig:A_re}, for very low-mass systems ($M_* \lesssim 10^5\, M_\odot$) the gas gravity can exceed the stellar gravity even in the optically thin regime.}, leading to $F_{\rm G} \simeq F_{\rm G, star}$. From eq. (\ref{eq:boost_factor}), the boost factor simplifies to
\begin{equation}
	A = \frac{\tau_{\rm ext} \sigma_{\rm pr}}{\sigma_{\rm T}} \propto N_{\rm H} \sigma_{\rm ex}\sigma_{\rm pr} \propto N_{\rm H} Z^2,
\end{equation}
where we have used eq. (\ref{eq:sigma_approx}) above. As $N_{\rm H} \simgt  10^{21}\, {\rm cm^{-2}}(Z/Z_\odot)^{-1}$, the system becomes optically thick. In this regime, the gravity due to the gas can still be neglected, $F_{\rm G} \simeq F_{\rm G, \rm star}$, and the boost factor becomes
\begin{equation}
	A = \frac{\sigma_{\rm pr}}{\sigma_{\rm T}} \propto Z.
\end{equation}

At even larger column densities, the system becomes highly optically thick and the gas gravity exceeds the stellar gravity, i.e., $F_{\rm G, gas} > F_{\rm G, star}$.  Such a threshold column density can be written as 
\begin{align}
 N_{\rm H, thres} &=  \frac{M_{\rm *}}{\frac{4}{3} \pi r_{\rm e}^3 m_{\rm p}} \, r_{\rm e} \notag
 &= 10^{23.5} \, {\rm cm^{-2}} \left(\frac{M_{\rm *}}{10^8 \,M_\odot}\right) \left(\frac{r_{\rm e}}{100 \,{\rm pc}}\right)^{-2}.
\end{align}

In this regime, the boost factor is given by
\begin{align}
 A &= \frac{F_{\rm G, star}}{F_{\rm G, star} + F_{\rm G, gas}} \frac{\sigma_{\rm pr}}{\sigma_{\rm T}} 
    = \left(1 + \frac{M_{\rm gas}}{M_*}\right)^{-1}  \frac{\sigma_{\rm pr}}{\sigma_{\rm T}}\notag  \\
    &= \left(1 + \frac{N_{\rm H}}{N_{\rm H, thres}}\right)^{-1} \frac{\sigma_{\rm pr}}{\sigma_{\rm T}}  \propto N_{\rm H}^{-1} Z.
\end{align}

To summarize, we can express the boost factor as a function of column density and metallicity in each regime: 
\begin{align}
   A = 
  \begin{cases}
    2.9 \times 10^{-18} N_{\rm H} (Z/Z_\odot)^2 \,\,\, &( N_{\rm H} < N_{\rm H ,\tau=1}) \\
    1.8\times 10^3 (Z/Z_\odot) \,\,\, &\left(N_{\rm H ,\tau=1} \leq N_{\rm H} < N_{\rm H, thres}\right) \\
    1.8 \times 10^3 \left(1+ \frac{N_{\rm H}}{N_{\rm H, thres}}\right)^{-1} (Z/Z_\odot) \, &(N_{\rm H} \geq  N_{\rm H, thres}), 
  \end{cases} \label{eq:boost_factor_in_func_NH_Z}
\end{align} 
where $N_{\rm H, \tau=1}$ is the threshold column density at which dusty gas becomes optically thick; 
\begin{equation}
N_{\rm H, \tau=1} = \frac{1}{\sigma_{\rm ext}}\approx 10^{21} \, {\rm cm^{-2}} \left(\frac{Z}{Z_\odot}\right)^{-1}. \label{eq:NH_opt}
\end{equation}

\section{Application to super-early galaxies} \label{sec:application_obsz10}
We now apply the boost factor calculation to JWST spectroscopically confirmed $z > 10$ galaxies, whose properties are shown in Table \ref{table:z10_obs}. Based on such data, we estimate gas column density, bolometric luminosity, and the boost factor and estimate the modified Eddington ratio for each galaxy as shown in Table \ref{table:z10_prediction}. 

We estimate $N_{\rm H}$ from the V-band attenuation, $A_{\rm V}$, as  
\begin{align}
    N_{\rm H} &= \frac{A_{\rm V}}{0.44 \sigma_{\rm ext, UV}} \notag \\
    &= 1.9 \times 10^{21} \, {\rm cm^{-2}} \left(\frac{A_{\rm V}}{0.1 \, {\rm mag}}\right) \left(\frac{Z}{0.1 Z_\odot}\right)^{-1}.  \label{eq:calc_NH_atten}
\end{align}
The coefficient 0.44 in eq. (\ref{eq:calc_NH_atten}) is valid for a screen geometry, i.e., $A_\lambda = 1.086 \tau_{\lambda}$ and a MW extinction curve for which, $A_{\rm UV} = 2.46 A_{\rm V}$\footnote{SMC extinction curve satisfies $A_{\rm UV} = 5.40 A_{\rm V}$.}.
The bolometric luminosity is calculated by BPASS using the observed values of $M_*$, age, and metallicity with an assumption of instantaneous star formation.  

We find that most of the observed $z > 10$ galaxies have the classical Eddington ratio $\lambda_{\rm E, mod} \simeq 10^{-3}$. The boost factors are in the range $1 \simlt A \simlt 270$. The large $A$ variation is induced by the metallicity scatter. In fact, as the estimated column densities are smaller than $N_{\rm H, \tau=1}$ (optically thin regime, see eq. (\ref{eq:NH_opt})), implying $A \propto N_{\rm H} Z^2$. The modified Eddington ratios are calculated by $\lambda_{\rm E, mod} = A\lambda_{\rm E, class}$. 

When assessing the onset of dusty outflow from $\lambda_{\rm E, mod}$, we take into account uncertainties in the observed physical properties. Most observed $z > 10$ galaxies lack detected emission lines, and the metallicities obtained from SED fitting are poorly constrained, often tending to be underestimated.
For instance, JADES-GS-z14-0 was initially reported to have \citep[$Z=0.014 Z_\odot$,][]{Carniani:2024, Helton:2024}. However, once the \OIII 88 \mum emission line was robustly detected by ALMA, such value has been updated to $Z \sim 0.2 \, Z_\odot$ \citep{Carniani:2024_z14_ALMA, Schouws:2024}. Also, the estimated column density can increase twice when adopting an SMC dust model. 

Since boost factors are highly sensitive to metallicity and column density (see eq. (\ref{eq:boost_factor_in_func_NH_Z})), the actual modified Eddington ratios might be larger than the ones in Table \ref{table:parameter} by a significant factor. Therefore, we lower the threshold for the modified Eddington ratio by a factor of 2.5, setting the new criterion to $\lambda_{\rm E, mod} > 0.4 (=\lambda_{\rm E, mod (thres)})$, and consider galaxies that meet this condition to be in the outflow phase. We find that three galaxies (GS-z12, CEERS2\_588, GN-z11) have $\lambda_{\rm E, mod}$ values above 0.4, with $\lambda_{\rm E, mod} = 1.635, 0.412, 0.457$, respectively.  

For the above three galaxies, we estimate the outflow velocity as follows \citet{Thompson:2015}. We assume a spherical, geometrically thin dusty gas shell of mass $M_{\rm sh}$, neglecting any further mass accumulation during the outflow. We also neglect the gas self-gravity, as its effect becomes negligible at larger radii.
The single-scattering momentum equation for the shell is 
\begin{align}
\dv{t}(M_{\rm sh}v) &= - G\frac{M_* M_{\rm sh}}{r^2} +f(\tau)\frac{L_{\rm bol}}{c}\label{eq:momentum_eq} \\
    f(\tau) &=1-e^{-\tau_{\rm ext}} + \tau_{\rm IR}. \label{eq:f_tau} 
\end{align}
Additionally, although the gravitational term from the central stellar component is explicitly included in eq. (\ref{eq:momentum_eq}), it is ultimately neglected under the assumption that radiation pressure dominates, as explained in Appendix \ref{appendix:terminal_velocity}.
The third term of eq. (\ref{eq:f_tau}) comes from the momentum exchange between IR radiation and the outflowing gas. It becomes effective when the gas is optically thick to IR. We calculate the optical depth to re-radiated IR photons for the four galaxies. The values for GS-z12, CEERS2\_558, and GN-z11 are $\tau_{\rm IR} =$ 0.31, 0.35, and 0.78, respectively, and all of the targets are optically thin to IR radiation\footnote{We adopt the dust mass absorption coefficient for IR photons of $\kappa_{\rm IR} = 117 \,{\rm cm^2 g^{-1}}$. The derivation is described in Appendix \ref{appendix:terminal_velocity}. We also adopt the gas mass as twice the stellar mass.}. Although the IR optical depth is $\sim$ 1000 times smaller than that of UV, IR radiation pressure may work during the initial evolutionary phase, before the outflow is fully launched.
Therefore, the terminal velocity is written as 
\begin{align}
    v_{\infty}  &= \left(\frac{2L}{c}\right)^{1/2} \left(\frac{\kappa_{\rm UV}}{\pi M_{\rm sh}}\right)^{1/4} \notag \\
      &= 153 \, {\rm km s^{-1}} \left(\frac{L}{10^{45}\, {\rm erg s^{-1}}}\right)^{1/2} \left(\frac{Z}{0.1 Z_\odot}\right)^{1/4}\left(\frac{M_{\rm sh}}{10^{9}\,M_\odot}\right)^{-1/4} .\label{eq:v_infty}
\end{align} 
The detailed calculation is described in Appendix \ref{appendix:terminal_velocity}. 

The shell mass, $M_{\rm sh}$, can be estimated from the observed $r_{\rm e}$ and $N_{\rm H}$. However, this corresponds to the shell mass at the observed epoch, not at the onset of the outflow. We therefore use the gas mass {\it just before the onset of the outflow} by assuming that it is twice the currently observed stellar mass, i.e., $f_{\rm gas} = 0.67$. The gas fraction value within $r_{\rm e}$ is derived in \citet{Ziparo:2023} assuming 10\% of a baryonic fraction within a halo. We approximate $M_{\rm sh} \sim M_{\rm gas}=f_{\rm gas}M_*/(1-f_{\rm gas})$ following \cite{Ferrara:2024_Lya}. Substituting the values of $L_{\rm bol}, M_*, Z$ in Table \ref{table:z10_obs} to eq. (\ref{eq:v_infty}), we obtain $v_{\infty} = 59, 97, 100\, {\rm km \,s^{-1}}$ and $V_{\rm circ}=70, 148, 149\,  {\rm km \,s^{-1}}$ for GS-z12, CEERS2\_588, and GN-z11, respectively. These values are 66-84\% of the host halos circular velocities\footnote{We derive the halo masses using the extrapolated stellar-to-halo mass relation in \citet{Behroozi:2013} as the same as \citet{Scholtz:2024} The obtained values are $\log_{10} M_{\rm halo}/M_\odot =$ 9.38, 10.4, 10.5 for for GS-z12, CEERS2\_588, and GN-z11, respectively.}, in good agreement with values inferred from observations \citep{Xu:2023, Carniani:2024_outflow}. 

We also calculate the dust-clearing timescale ($t_{\rm c}$) by using the outflow velocities for the three galaxies. Since the attenuation follows $A_{\rm V} \propto \tau_{\rm ext, UV} \propto r^{-2}$, we calculate $t_{\rm c}$ as the time it takes for the radius to expand by a factor of 10, thus decreasing $A_{\rm V}$ by 100 times. GS-z12, CEERS2\_588, and GN-z11 have $t_{\rm c} = 9 r_{\rm e}/v_{\infty} = 22$, 25, and 5.7 Myr, respectively. As $t_{\rm c}$ is much shorter than the dynamical timescale, this implies that outflows can quickly clear dust once they are launched. It is also interesting to compare the star formation and outflow timescales for these galaxies. We estimate the star formation duration timescale as the gas depletion timescale $t_{\rm dep} \sim M_{\rm gas}/{\rm SFR}$, and the outflow timescale as $t_{\rm out} = r_{\rm e}/v_\infty$. The obtained timescales for the GS-z12, CEERS2\_588, and GN-z11 are $t_{\rm dep} =$ 76 Myr, 100 Myr, 57 Myr, and $t_{\rm out} =$ 2.4 Myr, 2.8 Myr, 0.63 Myr, respectively. The outflow timescales are much smaller than the star formation duration timescales, implying that dusty outflows can contribute to the quenching of star formation. This is consistent with the result of \citep{Ferrara:2024_z14}, which predicts the time evolution of the observed $z \sim 14$ galaxy and finds that the SFR decreases by a factor of seven during the outflow phase. However, since the estimated outflow velocity is smaller than the escape velocity, the swept-up gas can contribute to an intermediate quenching (``mini-quenching''), and will be accumulated again to fuel further star formation.

The other 17 observed $z > 10$ galaxies have $\lambda_{\rm E, mod} \sim 10^{-3} - 0.3$, which indicates that they are not super-Eddington at the time of the observation. However, we cannot exclude that these galaxies already went through an outflow phase at an earlier evolutionary epoch. In the following we explore such a possibility.

\begin{table*}
    \centering
    \begin{tabular}{|l|l|l|l|l|l|l|l|l|l|l|}
    \hline
        &\multirow{2}{*}{Name} & \multirow{2}{*}{redshift} & \multirow{2}{*}{$\log_{10} \frac{M_*}{M_\odot}$} & $r_{\rm e}$ & age & SFR& \multirow{2}{*}{$\log_{10}\frac{Z}{Z_\odot}$} & \multirow{2}{*}{$A_{\rm v}$} & sSFR & \multirow{2}{*}{ref.} \\
     &     &     &     & [pc]   & [Myr] & $[M_\odot/{\rm yr}]$ &    &    & $[{\rm Gyr^{-1}}]$ &      \\
\hline\hline
        1 & JADES-GS-z14-0 & 14.32 & 8.7 & 260 & 10 & 22 & -1.5 & 0.31 & 45 & \citet{Carniani:2024} \\
        2 & JADES-GS-z14-1 & 13.9 & 8 & 160 & 10 & 2 & -1.1 & 0.2 & 18 &\citet{Carniani:2024} \\ 
        3 & GS-z13 & 13.2 & 7.7 & 59.3 & 40 & 1.412 & -1.9 & 0.022 & 28 & \citet{Hainline:2024} \\ 
        4 & UNCOVER-z13 & 13.08 & 8.13 & 309 & 66 & 1.28 & -1.57 & 0.0434 & 10 & \citet{Wang:2023} \\ 
        5 & JADES-GS-z13-1-LA & 13.05 & 7.74 & 61.6 & 21 & 0.16 & -2.52 & 0.04 & 3 & \citet{Witstok:2024} \\ 
        6 & GS-z12 & 12.48 & 7.64 & 146 & (10) & 1.15 & -0.70 & 0.052 & 26 & \citet{D'Eugenio:2023} \\ 
        7 & UNCOVER-z12 & 12.393 & 8.35 & 426 & 62 & 2.15 & -1.34 & 0.21 & 10 & \citet{Wang:2023} \\ 
        8 & GHZ2/GLASS-z12 & 12.34 & 9.05 & 34 & 28 & 5.2 & -1.34 & 0.04 & 5 & \citet{Castellano:2024} \\
        9 & Maisie & 11.42 & 8.6 & 360 & (10) & 2 & (-1) & 0.1 & 5 & \citet{Arrabal_Haro:2023} \\ 
        10 & GS-z11 & 11.122 & 8.3 & 118 & 158 & 1.445 & -1.9 & 0.043 & 7 & \citet{Hainline:2024} \\ 
        11 & CEERS2\_588 & 11.04 & 8.7 & 280 & (10) & 10 & (-1) & 0.2 & 20 & \citet{Harikane:2024} \\ 
        12 & GN-z11 & 10.603 & 8.73 & 64 & 10.23 & 18.78 & -0.92 & 0.17 & 35 & \citet{Bunker:2023} \\ 
        13 & GS-z10 & 10.38 & 7.9 & (100) & 31 & 1 & (-1) & (0.1) & 13 & \citet{Robertson:2023} \\ 
        14 & UNCOVER 37126 & 10.255 & 8.16 & (100) & 10 & 1.65 & (-1) & (0.1) & 11 & \citet{Atek:2023} \\ 
        15 & MACS0647-JD (A) & 10.17 & 7.5 & 70 & 5 & 2.4 & -0.70 & 0.01 & 76 & \citet{Hsiao:2023} \\ 
        16 & MACS0647-JD (B) & 10.17 & 7.7 & 20 & 50 & 1 & (-1) & 0.13 & 20 & \citet{Hsiao:2023}\\ 
        17 & CEERS2\_7929 (CEERS\_19996) & 10.1 & 8.66 & (100) & (10) & 1.9 & (-1) & (0.1) & 4 & \citet{Arraba_Haro:2023_ApJL} \\ 
        18 & UHZ-1 (UNCOVER 26185) & 10.073 & 8.14 & 592 & 64.9 & 1.25 & -0.70 & 0.08 & 9 & \citet{Goulding:2023} \\
        19 & CEERS\_35590 & 10.01 & 9.1 & 420 & (10) & 9 & (-1) & 0.1 & 8 & \citet{Arraba_Haro:2023_ApJL} \\ 
        20 & CEERS\_99715 & 9.97 & 9.5 & 580 & (10) & 6 & (-1) & 0.1 & 2 & \citet{Arraba_Haro:2023_ApJL} \\ \hline
    \end{tabular}
     \caption{Summary of the properties of $z > 10$ galaxies spectroscopically confirmed by JWST. {\bf Note.} (1) Object name. (2) Redshift determined by Lyman break or rest-frame optical lines. (3) Stellar mass. (4) Effective radius. (5) Mass-weighted stellar age. (6) SFR. (7) Metallicity. (8) Dust attenuation. (9) specific SFR. (10) References. Since some physical quantities are not referenced in the literature, we adopt typical values for galaxies at $z>10$ and they are enclosed in parentheses. } \label{table:z10_obs}
\end{table*}

\begin{table*}
    \centering
    \begin{tabular}{|l|l|l|l|l|l|l|l|l|}
    \hline
    & \multirow{2}{*}{Name} & \multirow{2}{*}{redshift} & $\log_{10} N_{\rm H}$ &$\log_{10} L_{\rm bol}$ &  $\log_{10} L_{\rm E}$& \multirow{2}{*}{$\lambda_{\rm E}$} & $\multirow{2}{*}{$A$}$ & \multirow{2}{*}{$\lambda_{\rm E, mod}$}\\ 
        &      &     & $[{\rm cm^{-2}}]$ 
        & [${\rm erg s^{-1}}$] & [${\rm erg s^{-1}}$] & & &  \\ \hline\hline
      1 &    JADES-GS-z14-0 &    14.32 &       22.3 &     44.6 & 46.8 & $6.41\times 10^{-3}$ &  29.5 &          0.189 \\
      2 &    JADES-GS-z14-1 &    13.90 &       21.7 &     43.9 & 46.1 & $6.32\times 10^{-3}$ &  53.7 &          0.339 \\
      3 &            GS-z13 &    13.20 &       21.5 &     42.9 & 45.8 & $1.23\times 10^{-3}$ &   1.1 &          0.001 \\
      4 &       UNCOVER-z13 &    13.08 &       21.5 &     43.1 & 46.2 & $8.01\times 10^{-4}$ &   3.1 &          0.003 \\
      5 & JADES-GS-z13-1-LA &    13.05 &       22.8 &     43.3 & 45.8 & $2.61\times 10^{-3}$ &   1.4 &          0.004 \\
      6 &            GS-z12 &    12.48 &       22.1 &     43.5 & 45.7 & $6.07\times 10^{-3}$ & 269.2 &          1.635 \\
      7 &       UNCOVER-z12 &    12.39 &       21.9 &     43.4 & 46.4 & $7.98\times 10^{-4}$ &  17.4 &          0.014 \\
      8 &              GHZ2 &    12.34 &       21.2 &     44.4 & 47.1 & $1.96\times 10^{-3}$ &   6.3 &          0.012 \\
      9 &            Maisie &    11.42 &       21.3 &     44.5 & 46.7 & $6.28\times 10^{-3}$ &  38.5 &          0.242 \\
     10 &            GS-z11 &    11.12 &       22.4 &     43.1 & 46.4 & $4.72\times 10^{-4}$ &   4.4 &          0.002 \\
     11 &        CEERS2\_588 &    11.04 &       21.6 &     44.6 & 46.8 & $6.28\times 10^{-3}$ &  65.7 &          0.412 \\
     12 &            GN-z11 &    10.60 &       21.4 &     44.6 & 46.8 & $6.23\times 10^{-3}$ &  73.3 &          0.457 \\
     13 &            GS-z10 &    10.38 &       21.3 &     43.2 & 46.0 & $1.49\times 10^{-3}$ &  27.8 &          0.041 \\
     14 &     UNCOVER 37126 &    10.26 &       21.3 &     44.1 & 46.3 & $6.28\times 10^{-3}$ &  39.1 &          0.246 \\
     15 &     MACS0647-JD\_A &    10.17 &       20.0 &     43.8 & 45.6 & $1.62\times 10^{-2}$ &  10.7 &          0.173 \\
     16 &     MACS0647-JD\_B &    10.17 &       21.4 &     42.8 & 45.8 & $9.17\times 10^{-4}$ &  31.4 &          0.029 \\
     17 &       CEERS2\_7929 &    10.10 &       21.3 &     44.6 & 46.8 & $6.28\times 10^{-3}$ &  39.2 &          0.246 \\
     18 &             UHZ-1 &    10.07 &       24.0 &     43.1 & 46.2 & $7.24\times 10^{-4}$ &   2.8 &          0.002 \\
     19 &       CEERS\_35590 &    10.01 &       21.3 &     45.0 & 47.2 & $6.28\times 10^{-3}$ &  38.9 &          0.244 \\
     20 &       CEERS\_99715 &     9.97 &       21.3 &     45.4 & 47.6 & $6.28\times 10^{-3}$ &  39.0 &          0.245 \\
     \hline
    \end{tabular}
    \caption{{\bf Notes.} (1) Object name. (2) Redshift. (3) The gas column density calculated using eq. (\ref{eq:calc_NH_atten}). (4) Bolometric luminosity calculated from BPASS using the observed values of $M_*$, age, and metallicity from Table \ref{table:z10_obs}. (5) Classical Eddington luminosity based on $M_*$ from Table \ref{table:z10_obs}. (6) Classical Eddington ratio, $\lambda_{\rm E}\equiv L_{\rm bol}/L_E$. (7) Boost factor. (8) Modified Eddington ratio,, $\lambda_{\rm E, mod} = A\lambda_{\rm E}$. For the column density of UHZ-1, we adopt the value reported by \citet{Goulding:2023}, which indicates that this object is Compton-thick to X-rays.}
    \label{table:z10_prediction}
\end{table*}

\subsection{A bursty phase prior to observation?}\label{subsec:application_SFH}
\citet{Ferrara:2024_z14} develop an analytical model for the star formation history (SFH) of high-redshift galaxies, and applied it to GS-z14-0 located at $z=14.32$. Their model reproduces the physical properties of GS-z14-0 when the model galaxy becomes super-Eddington 66 Myrs before the observation and a dusty outflow continues for 40 Myrs. We, therefore, investigate if the observed galaxies have experienced the dusty outflow phase a few $\times$ 10 Myr before the observations.
We consider the delayed-$\tau$ star formation model according to which the SFR exponentially declines with time after an initial burst occurring at $t = t_{\rm SF}$, 
\begin{equation}
    {\rm SFR} (t) \propto 
    \begin{cases}
        0 \,\,(t < t_{\rm SF}) \\
        (t-t_{\rm SF})\exp \left(-\frac{t-t_{\rm SF}}{\tau}\right) \,\,(t > t_{\rm SF}). 
    \end{cases}
\end{equation}
The corresponding evolution of the stellar mass is 
\begin{align}
    M_*(t) &= 
    \begin{cases}
        0 \,\,(t < t_{\rm SF}) \\
        M_0 \tau \left[\tau- (t-t_{\rm SF} + \tau)e^{-(t-t_{\rm SF})/\tau}  \right] (t > t_{\rm SF}),
    \end{cases} \\
    M_0 &= \frac{M_*(t_{\rm obs})}{\tau \left[\tau - (\tau_{\rm age}+ \tau)e^{-\tau_{\rm age}/\tau} \right]}, 
\end{align}
where $\tau_{\rm age}= t_{\rm obs}- t_{\rm SF}$ is roughly equal to mass-weighted stellar age obtained from observation (Table \ref{table:z10_obs}).  The constant $M_0$ is obtained by imposing $\int {\rm SFR}(t) {\rm d}t = M_*(t_{\rm obs})$. For typical star formation timescale $\tau$, we assume $\tau = 2\, {\rm Myr}$, which is the free-fall time at $n_{\rm H}\sim 10^3\, {\rm cm^{-3}}$. We assume that galaxy sizes do not change largely and set $r_{\rm e}(t) = r_{\rm e}(t_{\rm obs})$. For metallicity, we set 
\begin{equation}
    Z(t) = Z(t_{\rm obs}) \left(\frac{M_*(t)}{M_*(t_{\rm obs})}\right)^x, 
\end{equation}
where the power $x = 0.3-0.5$ is the slope of the mass-metallicity relation inferred from recent observations and theoretical studies at $z \gtrsim 9$ \citep[e.g.][]{Nakajima:2023, Nakazato:2023, Langan:2020, Tsuna:2023};  we adopt $x=0.5$. We assume that the initial gas mass is twice the observed stellar mass, i.e., $M_{\rm gas}(t)= 2M_*(t_{\rm obs})- M_*(t)$. The column density at time $t$ at the outer radius is calculated as $N_{\rm H}(t) = n_{\rm H} × r_{\rm e}$. Similarly, the corresponding optical depth at time t is calculated as $\tau_{\rm ext}(t) = N_{\rm H}(t) \sigma_{\rm ext}$. As mentioned in Section \ref{subsubsec:modified_eddington_luminosity}, our definition of $\sigma_{\rm ext}$ explicitly includes the dust-to-gas ratio $\mathcal{D}$.
The bolometric luminosity is derived as $L_{\rm bol} = f_{\rm bol} L_{\rm UV}$, where $f_{\rm bol}=2$ \citep{Fiore:2023}.  is a bolometric correction, and $L_{\rm UV}$ is the intrinsic UV luminosity computed as $L_{\rm UV} = \mathcal{K}_{\rm UV} {\rm SFR}$. Here $\mathcal{K}_{\rm UV}$ is a conversion factor and its value is chosen to match the one used by the ALMA REBELS survey $\mathcal{K}_{\rm UV} = 0.587\times 10^{10}\, L_\odot/M_\odot {\rm yr}^{-1}$ \citep{Bouwens:2022}.

\begin{figure*}
    \centering
    \includegraphics[width = \linewidth, trim=0 12 0 0, clip]{\figdir/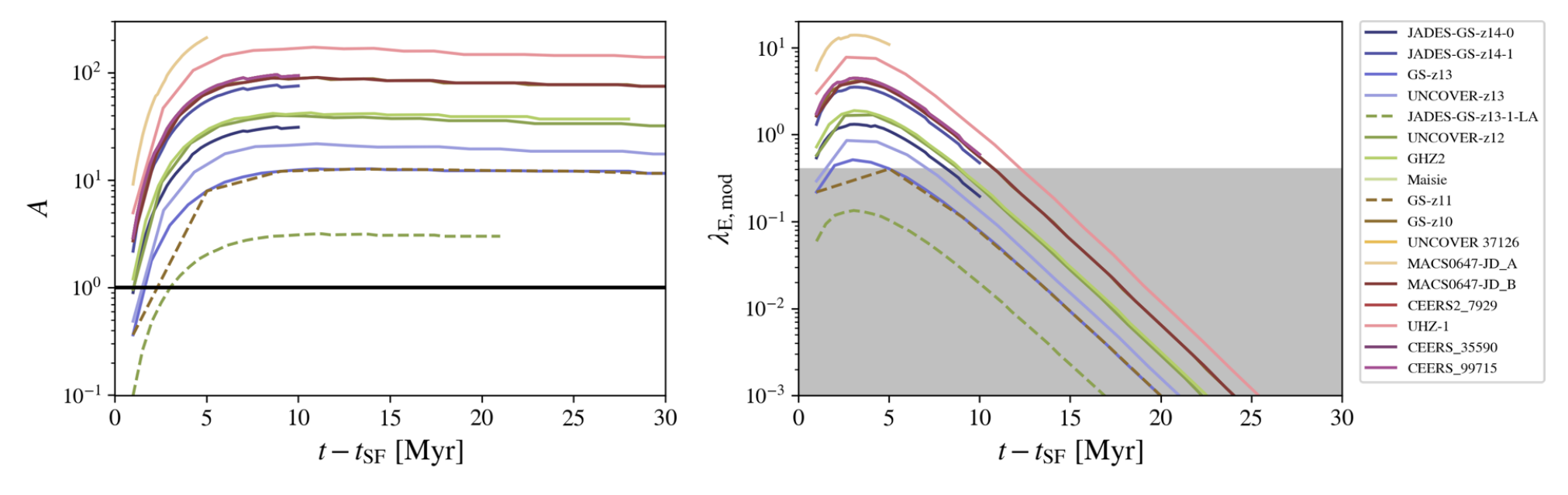}
    \vspace{-0.6cm}
    \caption{Time evolution of boost factor $A$ (left) and the modified Eddington ratios $\lambda_{\rm E, mod}$ (right). The gray regions from $\lambda_{\rm E, mod} < 0.4$ indicate the rough boundary where dusty outflows are not onset. We plot the results of JADES-GS-z13-LA and GS-z11 as dashed lines to show they have not experienced outflow phases.
}
    \label{fig:B_lambda_mod_SFH}
\end{figure*}

Figure \ref{fig:B_lambda_mod_SFH} shows the time evolution of $A$  (left) and modified Eddington ratios (right) for the other 17 galaxies. We find that 15 out of 17 galaxies have a phase where $\lambda_{\rm E, mod} \gtrsim 0.4$ and therefore should have experienced outflows before the observation. The time evolution of  $\lambda_{\rm E, mod}$ closely tracks that of the SFR. It reaches a peak at $t-t_{\rm SF} \sim \tau=2\, {\rm Myr}$. For the 15 objects exceeding the threshold, we calculate the clearing timescale, $t_{\rm c}$, as defined in Section \ref{sec:application_obsz10} by adopting the physical properties at $t- t_{\rm SF} = \tau$. We find that 10 out of 15 galaxies have $(\tau_{\rm age} -\tau > t_{\rm c})$, which indicates that outflows can expel the produced dust and effectively decrease $A_V$ by the time of observation.\\
\indent
We note that in Figure \ref{fig:B_lambda_mod_SFH} MACS0647-JD(A) has $\lambda_{\rm E,mod} > 1$ at the observed epoch, even though it is shown to have $\lambda_{\rm E,mod} < 1$ when using the observed physical properties in Table \ref{table:z10_prediction}. This discrepancy arises from our assumption of the initial gas mass. For MACS0647-JD(A) this gives $M_{\rm gas} \approx 10^{7.5}\,M_\odot$, corresponding to $n_{\rm gas} \approx 890\,{\rm cm^{-3}}$ and $\log N_{\rm H} \approx 23.3\,{\rm cm^{-2}}$. By contrast, the observed column density is only $\log N_{\rm H} \approx 20.0\,{\rm cm^{-2}}$, which is below $N_{\rm H}(\tau=1)$ and thus in the optically thin regime. The overestimation of the gas mass in our SFH model for MACS0647-JD(A) results in a boost factor $A$ that is larger than the one derived directly from the observed properties.

We also calculate the time evolution of the boost factor $A$ and the modified Eddington ratio for the three galaxies assumed to be in the outflow phase: GS-z12, CEERS2\_588, and GN-z11. The boost factors at the observed time obtained from the SFH models are $A_{\rm SFH}(t_{\rm obs}) =$ 184, 94, and 112 for GS-z12, CEERS2\_588, and GN-z11, respectively. These values differ from those inferred directly from the observed quantities by a factor of $\lesssim 1.5$. The corresponding $\lambda_{\rm mod}$ values therefore scale in the same way, but still above the threshold value ($\lambda_{\rm mod, thres}$).

\subsection{Caveats}
Our model contains some approximations that are worth discussing.  First, we assume a spherical gas distribution and a concentrated stellar distribution as a point source. {The point-like approximation can be applicable to compact galaxies observed at $z > 10$. This approximation becomes more accurate as the outflow develops and the gas is spread on scales much larger than $r_{\rm e}$.} However, massive galaxies in the Epoch of Reionization ($z\sim 6-8$) with stellar masses of $M_* \gtrsim 10^9 \, M_\odot$ are known to have more complex morphologies, such as clumpy structures or extended disks, as revealed by both observations and simulations \citep[e.g.][]{Chen:2023, Hainline:2024, Hainline:2024_clumpy, Rowland:2024, Harikane:2024_bright_end, Nakazato:2024, Kohandel:2024}. 

In our modeling, we also assume a homogeneous gas metallicity within the effective radius ($R_{\rm eff}$). This approximation is justified by the very compact nature of the galaxies under study ($R_{\rm eff}\sim 100 \,{\rm pc}$). Observational constraints on metallicity gradients in high-redshift galaxies have so far been limited to massive, extended systems at $z \sim 6-8$ using JWST/NIRSpec IFU and GRISM data \citep[e.g.,][]{Venturi:2024, Vallini:2024, Arribas:2024}. These works report nearly flat gradients (e.g., –0.03 ${\rm dex\, kpc^{-1}}$), although some simulations predict somewhat steeper gradients \citep[e.g., –0.15 ${\rm dex\, kpc^{-1}}$, ][]{Garcia:2025}. Given the small sizes of our target galaxies, we therefore consider the homogeneous metallicity assumption to be reasonable.

Our model assumes smooth and uniform gas distributions. In reality, clumpy or turbulent structures will broaden the distribution of column densities, leading to significant variation of the boost factor $A$ along different sightlines. Because $A$ scales approximately inversely with column density, low-column sightlines can reach super-Eddington conditions even if the average Eddington ratio is below unity. Turbulence may therefore contribute to anisotropic radiation-driven outflows \citep{Thompson:2016}, implying that our results neglecting turbulence are conservative. In addition, swept-up shells may become self-gravitating and fragment, as in the ``collect-and-collapse'' scenario studied around \HII regions \citep[e.g.,][]{Whitworth:1994, Hosokawa_Inutsuka:2006, Dale:2007, Kim:2018}. For our systems, the dust-clearing timescale ($t_{\rm c} \sim 5-25 \, {\rm Myr}$) exceeds the free-fall time when $n_{\rm gas} \gtrsim 5\, {\rm cm^{-3}}$, suggesting that fragmentation may occur. Such fragmentation would produce clumps, altering the column density distribution and potentially leading to anisotropic outflows, while also providing additional radiation sources from newly formed stars.

In this work, we have focused exclusively on radiation pressure as a feedback mechanism. Other processes such as supernova explosions may also contribute, but the efficiency of SN-driven outflows is expected to be very low in the dense environments of compact galaxies at $z > 10$. The kinetic energy of SNe is quickly lost to radiative cooling on timescales of a few $\times$ 100 yr \citep{Terlevich:1992, Pizzati:2020}, and simulations show that SN-driven shocks often fail to escape and instead result in fallback \citep{Bassini:2023, Nath:2023, Ferrara:2024_Lya, Ferrara:2024_z14}. Thus radiation pressure is likely the dominant feedback channel at these epochs. A full assessment of multi-channel feedback will require future RHD cosmological simulations. However, such calculations demand both large volumes ($L_{\rm box} \sim 100 \,{\rm cMpc}$) and sub-pc resolution, which are currently computationally prohibitive. Recent progress has been made with smaller-scale radiation-hydrodynamical (RHD) simulations of stellar cluster formation \citep[e.g.,][]{Krumholz_Thompson:2013, Fukushima:2020, Menon:2023, Menon:2024, Menon:2025}, which primarily model radiation pressure and often omit winds and SNe feedback due to their limited timescales.

Also, in realistic galaxies, gas and stars are well-mixed as seen in simulations \citep[e.g.][]{Ceverino:2024, Pallottini:2022}. In this case, the radiative pressure from stars can become more complex, where it might not always work as an outward pressure but as an inward pressure. We plan to treat this complex geometry in future work by performing RHD calculations within zoom-in cosmological simulations. 

Furthermore, we assume that dust and gas are tightly dynamically coupled.  To check if this is a reasonable assumption, we compare the mean free path for grain-atom collisions and galaxy size based on \citet{Murray:2005}. The mean free path is $\lambda = (n_{\rm H} \sigma_{\rm dg})^{-1}$, where $\sigma_{\rm dg}$ is the dust-gas scattering cross section. Since we require that order of unity of the dust momentum imparted to the gas, the quantity of interest is $\lambda_{\rm M} = \lambda (m_{\rm D}/m_{\rm p})$, where $m_{\rm D}$ is the mass of an individual spherical dust grain of radius $a$. Considering the geometric cross section, we estimate 
\begin{align}
 \lambda_{\rm M} &= \frac{1}{n_{\rm H}a^2 \pi}\frac{m_{\rm D}}{m_{\rm p}} \notag \\
     &= 7.8 \, {\rm pc} \left(\frac{a}{0.1\, {\rm \mu m}}\right)\left(\frac{n_{\rm H}}{1\, {\rm cm^{-3}}}\right)^{-1}  \left(\frac{\rho_{\rm d}}{3\, {\rm g cm^{-3}}}\right), 
\end{align}
where $\rho_{\rm d}$ is the mass density of dust grains.

Therefore, by comparing $\lambda_{\rm M}$ with the typical size of observed $z >10$ galaxies ($\sim 100\,{\rm pc}$), we find $\lambda_{\rm M} \ll r_{\rm e}$. Hence, the dust is efficiently coupled with the gas, transferring the radiation field momentum to it.
However, at low column densities  
in Figure \ref{fig:A_colormap}), it is possible to have $\lambda_{\rm M} > r_{\rm e}$ and thus for dust and gas to be decoupled. This scenario might allow the outflow to sweep away only the dust from the galaxy while leaving enough gas behind to support continued star formation.

\section{Summary} \label{sec:conclusion}
We have calculated the modified Eddington luminosity in a dusty medium as a function of 8 galaxy parameters: stellar mass, gas fraction, size, gas distribution, dust properties, dust-to-gas ratio, stellar age, and metallicity. We have introduced the boost factor $A \equiv L_{\rm E}/L_{\rm E, mod}$, which represents the ratio between the classical Eddington luminosity and the modified Eddington luminosity in a dusty medium.

Unlike the classical Eddington luminosity, which only considers Thomson scattering, the modified Eddington condition accounts for both the absorption and scattering of photons (primarily UV) by dust, which lowers the modified Eddington luminosity and relaxes the conditions for super-Eddington outflows to occur. The concepts of the boost factor and the modified Eddington luminosity have been previously introduced mainly in the context of AGN-driven dusty outflows \citep[e.g.,][]{Fabian:2006, Fabian:2008, Thompson:2015, Ishibashi:2015, Costa:2018_isolated, Costa:2018_cosmological}. Here we apply this framework to stellar radiation-driven systems, in particular to newly observed galaxies at $z > 10$. The main findings are:
\begin{itemize}
    \item The boost factor $A$ primarily depends on gas metallicity, $Z$, and column density, $N_{\rm H}$. This is because the dust-to-gas mass ratio scales with metallicity, and the column density mainly reflects the optical depth of dusty gas and the contribution of gas gravity. Expressions for A in different optical depth regimes are given simple, but accurate expression for the boost factor is as expressed in eq. (\ref{eq:boost_factor_in_func_NH_Z}).  For $\tau \approx 1$, it is $A = 1800 (Z/Z_\odot)$. 
    \item We apply the above results to the 20 galaxies currently observed at $z \gtrsim 10$ and derive both $A$ and the modified Eddington ratio $\lambda_{\rm E, mod}$. Three galaxies (GS-z12, CEERS2\_588, and GN-z11) are likely in the outflow phase, with the outflow velocities of 60-100 ${\rm km\ s^{-1}}$.
    \item For the other $z > 10$ galaxies, we model their star formation histories (SFH) to check whether they have experienced an outflow phase before the observation epoch. Assuming a delayed-$\tau$ model, we find that 15/17 galaxies had $\lambda_{\rm E, mod}> \lambda_{\rm E, mod (thres)} (= 0.4)$  before the observation. We estimate the dust-clearing timescale ($t_{\rm c}$) required to reduce $A_{\rm V}$ by a factor of 100 and find that 10/15 galaxies have a clearing time much shorter than their stellar ages. This implies that outflows have effectively expelled the dust prior to observation.
\end{itemize}

The above results propose the possibility that radiation-driven outflows can be triggered during the early phases of galaxy evolution probed by JWST at high redshifts. Such outflows might contribute to the observed UV brightness of these systems and to regulating their evolution.
The investigation of such issues will require future dedicated theoretical and numerical modeling. On the observational side, ALMA may identify outflows with its high velocity resolution ($\Delta v \sim 10-15\, {\rm km s^{-1}}$), compared to JWST/NIRSpec ($\Delta v \sim 50\, {\rm km/s}$ at $R \sim 2700$). Just recently, the ALMA Large Program of Cycle 12, PHOENIX (PI: Schouws), has been accepted, targeting \OIII 88\mum lines from 10 galaxies at $z = 8-15$. These observations will provide direct tests of whether the observed galaxies are in an outflow phase.

\section*{Acknowledgements}
We are grateful to an anonymous referee for
valuable comments that have greatly improved
the paper.
We thank N. Yoshida, Y. Harikane, T. Hosokawa, M. Ouchi, K. Omukai, H. Yajima, H. Fukushima for useful discussions. YN acknowledges funding from JSPS KAKENHI Grant Number 23KJ0728, JSPS International Leading Research 23K20035, and JSR fellowship. AF acknowledges support from the ERC Advanced Grant INTERSTELLAR H2020/740120.  Partial support from the Carl Friedrich von Siemens-Forschungspreis der Alexander von Humboldt-Stiftung Research Award is kindly acknowledged. This research was supported in part by grant NSF PHY-2309135 to the Kavli Institute for Theoretical Physics (KITP). The Flatiron Institute is a division of the Simons Foundation.
This work made use of v2.3 of the Binary Population and Spectral Synthesis (BPASS) models as described in \citet{Byrne:2022} and \citet{Stanway:2018}.

 \section*{Data Availability}
 All data underlying this article are available on reasonable request to the corresponding author.



\bibliographystyle{mnras}
\bibliography{dusty_outflow} 



\appendix
\section{Calculation of terminal velocity} \label{appendix:terminal_velocity}
Here we derive the terminal velocity $v_{\infty}$ based on \citet{Thompson:2015}. We first assume that the mass of the shell is constant as a function of radius (i.e., the shell expands into vacuum.)
The general expression for momentum conservation for a thin shell of mass $M_{\rm sh}$ is 
\begin{align}
\dv{t}(M_{\rm sh}v) &= - G\frac{M_* M_{\rm sh}}{r^2} +(1-e^{-\tau_{\rm UV}} + \tau_{\rm IR})\frac{L_{\rm bol}}{c}\\
\therefore \, M_{\rm sh} v\dv{v}{r}& = - G\frac{M_* M_{\rm sh}}{r^2} +(1-e^{-\tau_{\rm UV}} + \tau_{\rm IR})\frac{L_{\rm bol}}{c} \label{eq:momentum_eq_appendix}
\end{align} 
where $\tau_{\rm UV, IR}$ is the optical depth of the shell to UV(IR) photons; 
\begin{equation} 
\tau_{\rm UV, IR} = \frac{\kappa_{\rm UV, IR}M_{\rm sh}}{4\pi r^2} .
\end{equation}
Here $\kappa_{\rm UV, IR}$ is the dust mass absorption coefficient for UV (IR) photons. Adopting a  MW extinction with a dust-to-gas ratio $\mathcal{D}_{\rm MW} = 1/165$, the value at 1500 \AA \, is 
$\kappa_{\rm UV} =  \kappa_{\rm 1500}= 1.26\times 10^5 \, {\rm cm^2 g^{-1}}(\mathcal{D}/\mathcal{D}_{\rm MW})$ \citep{Ferrara:2022}. For the IR coefficient, $\kappa_{\rm IR}(\nu) = \kappa_{158}(\nu/\nu_{158})^{\beta_{\rm d}}$, is pivoted at wavelength $\lambda = 158 $\mum.  For a MW curve, $\kappa_{158} = 10.41 \,{\rm cm^2 g^{-1}}$, and $\beta_{\rm d} = 2.03$ \citep{Weingartner:2001}. 
 Dust temperature ($T_{\rm d}$) increases as redshift is higher, and reaches values of 50-70 K at $z\simeq 10$ \citep[e.g.,][]{Sommovigo:2022}. We adopt $T_{\rm d} = 60\, {\rm K}$ as a fiducial value and compute $\kappa_{\rm IR}$ at the peak wavelength of the gray-body spectrum, $\lambda_{\rm p} =0.29/T_{\rm d} = 48 \,$\mum, obtaining $\kappa_{\rm IR}(\nu_{\rm p}) = 117\, {\rm cm^2 g^{-1}}(\mathcal{D}/\mathcal{D}_{\rm MW})$. Therefore, the opacity for IR photons is over 1000 times lower than that for UV photons.

We introduce a characteristic radius, $R_{\rm UV}$, at which the shell become optically thin to UV photons; 
\begin{equation}
    R_{\rm UV} = \left(\frac{\kappa_{\rm UV}M_{\rm sh}}{4\pi} \right)^{1/2} 
                    =  14.5 \,{\rm kpc} \left(\frac{M_{\rm gas}}{10^9 \, M_\odot}\right)^{1/2} \left(\frac{Z}{0.1 Z_\odot}\right)^{1/2}. \label{eq:r_tau1} 
\end{equation}

Integrating eq. (\ref{eq:momentum_eq_appendix}) from the initial distance $R_{\rm 0}$ to $R_{\rm UV}$, the left hand side of eq. (\ref{eq:momentum_eq_appendix}) becomes
\begin{equation} 
 \int^{R_{\rm UV}}_{R_0} M_{\rm sh} v \dv{v}{r} {\rm d}r = \frac{M_{\rm sh}}{2} (v_{\rm UV}^2 - v_0^2). 
\end{equation}
Here $v_{\rm UV}, v_0$ are the velocities at $R_{\rm UV}, R_0$. 
The right-hand side of eq. (\ref{eq:momentum_eq_appendix}) becomes
\begin{align}
 &\int^{R_{\rm UV}}_{R_0} -\frac{GMM_{\rm sh}}{r^2} {\rm d}r + \int^{R_{\rm UV}}_{R_0} (1+\tau_{\rm IR}) \frac{L}{c} {\rm d} r  \notag \\
&= GMM_{\rm sh}\left(\frac{1}{R_{\rm UV}} - \frac{1}{R_0}\right) + \frac{L}{c} \left[R_{\rm UV} - R_0 + \frac{\kappa_{\rm IR}M_{\rm sh}}{4\pi}\left(\frac{1}{R_{\rm 0}} - \frac{1}{R_{\rm UV}}\right)\right] \notag \\
&= \left[-\frac{GMM_{\rm sh}}{R_0} + \frac{L}{c}\left(R_{\rm UV} + \frac{\kappa_{\rm IR}M_{\rm sh}}{4\pi R_0}\right) \right] \left(1- \frac{R_0}{R_{\rm UV}}\right). \label{eq:integration_UV_optically_thick} 
\end{align} 

We also integrate eq. (\ref{eq:momentum_eq_appendix}) in the regions at $R_{\rm UV} \leq r \leq \infty$, where the shell is optically thin to UV photons. The left-hand side of eq. (\ref{eq:momentum_eq_appendix}) becomes
\begin{equation}
\frac{M_{\rm sh}}{2} (v_{\infty}^2 - v_{\rm UV}^2) .
\end{equation}
The right-hand side of eq. (\ref{eq:momentum_eq_appendix}) becomes
\begin{equation}
 -\frac{GMM_{\rm sh}}{R_{\rm UV}} + \int^\infty_{R_{\rm UV}} \frac{L\tau_{\rm UV}}{c} {\rm d} r 
= -\frac{GMM_{\rm sh}}{R_{\rm UV}} + \frac{L}{c} \frac{\kappa_{\rm UV}M_{\rm sh}}{4\pi R_{\rm UV}}.
\end{equation}

Therefore, integrating $0 \leq r \leq \infty$, the left hand side becomes
\begin{equation}
M_{\rm sh}\frac{v_{\infty}^2}{2} -M_{\rm sh}\frac{v_0}{2} \sim M_{\rm sh}\frac{v_{\infty}^2}{2}.
\end{equation}
We take the zero initial shell velocity ($v_0 \sim 0$), 
For the right hand side, taking $R_{\rm 0} \ll R_{\rm UV}$ and $\kappa_{\rm IR} \ll \kappa_{\rm UV}$, 
\begin{align}
&\left[-\frac{GMM_{\rm sh}}{R_0} + \frac{L}{c}\left(R_{\rm UV} + \frac{\kappa_{\rm IR}M_{\rm sh}}{4\pi R_0}\right) \right] \left(1- \frac{R_0}{R_{\rm UV}}\right) + \frac{M_{\rm sh}}{R_{\rm UV}} \left(\frac{\kappa_{\rm UV}L}{4\pi c}- GM\right) \notag \\
& \simeq -\frac{GMM_{\rm sh}}{R_0} \left(1 + \frac{R_0}{R_{\rm UV}}\right) + \frac{L R_{\rm UV}}{c} +
\frac{M_{\rm sh}L}{4\pi c}\frac{\kappa_{\rm UV}}{R_{\rm UV}}\left(1 + \frac{\kappa_{\rm IR}}{\kappa_{\rm UV}}{R_0}{R_{\rm UV}}\right) \notag \\
&\simeq -\frac{GMM_{\rm sh}}{R_0} + \frac{LR_{\rm UV}}{c} + \frac{LM_{\rm sh}}{4\pi c}\frac{\kappa_{\rm UV}}{R_{\rm UV}}. \label{eq:right_hand_side}
\end{align}
Assuming that the radiative pressure term is significantly large and the gravity term in eq.(\ref{eq:right_hand_side}) can be neglected, we obtain
\begin{align}
\frac{v_{\infty}^2}{2} &=  \frac{LR_{\rm UV}}{cM_{\rm sh}} + \frac{L}{4\pi c}\frac{\kappa_{\rm UV}}{R_{\rm UV}} = \frac{L}{c} \left(\frac{\kappa_{\rm UV}}{\pi M_{\rm sh}}\right)^{1/2} \\
\therefore v_\infty &= \left(\frac{2L}{c}\right)^{1/2} \left(\frac{\kappa_{\rm UV}}{\pi M_{\rm sh}}\right)^{1/4} \notag \\
&= 153 \, {\rm km s^{-1}} \left(\frac{L}{10^{45}\, {\rm erg s^{-1}}}\right)^{1/2} \left(\frac{Z}{0.1 Z_\odot}\right)^{1/4}\left(\frac{M_{\rm sh}}{10^{9}\,M_\odot}\right)^{-1/4} .
\end{align}

\bsp	
\label{lastpage}
\end{document}